\documentstyle[aps,floats,epsfig,twocolumn]{revtex}

\begin{document}
\title{Mechanisms of Spontaneous Current Generation in an Inhomogeneous $d$-Wave
Superconductor}
\author{M.H.S. Amin$^{a}$, A.N. Omelyanchouk$^{b}$, and A.M. Zagoskin$^{a,c}$}
\address{$^{a}$D-Wave Systems Inc., 320-1985 W. Broadway, Vancouver, B.C., V6J 4Y3,
Canada \\ $^{b}$B.I.Verkin Institute for Low Temperature Physics
and Engineering, \\ Ukrainian National Academy of Sciences, Lenin
Ave. 47, Kharkov 310164, Ukraine \\ $^{c}$Physics and Astronomy
Dept., The University of British Columbia,\\ 6224 Agricultural
Rd., Vancouver, B.C., V6T 1Z1, Canada.}
\date{\today}

\address{~
\parbox{14cm}{\rm
\medskip
A boundary between two $d$-wave superconductors or an $s$-wave
and a $d$ -wave superconductor generally breaks time-reversal
symmetry and can generate spontaneous currents due to proximity
effect. On the other hand, surfaces and interfaces in $d$-wave
superconductors can produce localized current-carrying states by
supporting the ${\cal T}$-breaking combination of dominant and
subdominant order parameters. We investigate spontaneous currents
in the presence of both mechanisms and show that at low
temperature, counter-intuitively, the subdominant coupling {\em
decreases} the amplitude of the spontaneous current due to
proximity effect. Superscreening of spontaneous currents is
demonstrated to be present in any $ d$-$d$ (but not $s$-$d$)
junction and surface with $d+id^{\prime}$ order parameter
symmetry. We show that this supercreening is the result of
contributions from the local magnetic moment of the condensate to
the spontaneous current. }} \maketitle
%\pacs{}

The time-reversal symmetry (${\cal T}$) breaking on surfaces and
interfaces of superconductors with $d$-wave orbital pairing has
been intensively investigated in the last years both in theory
and experiment
\cite{sigrist,FogelstromYip,LSW,Iguchi,covington,TafuriKirtley,Yip,Huck}.
Several mechanisms of ${\cal T}$-breaking have been proposed,
which fall in two categories: appearance of subdominant order
parameter and proximity effect\cite{FogelstromYip,LSW}.

In the first case the surface or interface suppresses the
dominant order parameter ($d_{x^{2}-y^{2}}$ in YBCO
\cite{Iguchi}). If the pairing interaction in other channels is
nonzero, the subdominant order parameter will be formed below the
corresponding, smaller critical temperature $T_{c2}$
\cite{LANDAFSHIC}. The combination of the two order parameters
with complex coefficients breaks the ${\cal T}$-symmetry
\cite{sigrist} and leads to spontaneous surface currents and
magnetic fluxes. Usually $d_{x^{2}-y^{2}}\pm is$ or
$d_{x^{2}-y^{2}}\pm id_{xy}$ combinations are predicted. Recent
observations of zero bias peak splitting in surface tunneling
experiments \cite{covington} and spontaneous fractional flux
(0.1-0.2 $\Phi _{0}$) near the ``green phase'' inclusions in YBCO
films \cite{TafuriKirtley} agree with this picture.

The other possibility arises in a junction between two $d$-wave
superconductors with different orientations of the order
parameter \cite{Yip}. In this case the two order parameters
necessary to form a ${\cal T}$-breaking state, $d_{1,2},$ are
supplied by the bulk superconductors. The equilibrium phase
difference across the boundary, $\phi _{0},$ is generally neither
0 nor $\pi ,$ and therefore the states with $d_{1}+e^{\pm i\phi
_{0}}d_{2}$ orderings are degenerate and may support spontaneous
currents. The same mechanism applies in case of a boundary
between an $s$- and a $d$-wave superconductor\cite{Huck}.

In order to investigate the interplay of both mechanisms, in this
letter we consider $d$-$d$ and $s$-$d$ interfaces as well as
(110)-surface of a $d$-wave superconductor. We will see that
generally the spontaneous currents due to proximity effect are
suppressed by the existence of subdominant order parameter. There
is also an important distinction between the $d$-$d$ and $s$-$d$
cases: In the former case the superconductor may have local
orbital and magnetic moments, contributing to the non-dissipative
current. In the latter case such a contribution is absent. Our
results indicate that in a clean $d$-$d$ junction all of the
spontaneous current can be attributed to this ``molecular
currents'' mechanism. We also show that this effect leads to
``superscreening'' of spontaneous currents in $d$-$d$ junctions
(i.e. to the existence of counter-currents independent of the
Meissner effect).

We use the standard approach based on quasiclassical Eilenberger equations
for Green's functions integrated over energy \cite{Eilenberger}
\begin{equation}
{\bf v}_{F}\cdot \nabla \widehat{G}_{\omega }+[\omega
\widehat{\tau }_{3}+ \widehat{\Delta },\widehat{G}_{\omega }]=0,
\label{eil}
\end{equation}
where $\omega $ is the Matsubara frequency and
\[
\widehat{G}_{\omega }({\bf v}_{F},{\bf r})=\left(
\begin{array}{cc}
g_{\omega } & f_{\omega } \\
f_{\omega }^{\dagger } & -g_{\omega }
\end{array}
\right) ,\quad \widehat{\Delta }({\bf v}_{F},{\bf r})=\left(
\begin{array}{cc}
0 & \Delta \\
\Delta ^{\dagger } & 0
\end{array}
\right) .
\]
Here $\widehat{G}_{\omega }$ is the matrix Green's function and
$\Delta $ is the superconducting order parameter. They both are
functions of Fermi velocity ${\bf v}_{F}$ and position ${\bf r}$.
We also need to satisfy the normalization condition $g_{\omega
}=\sqrt{1-f_{\omega }f_{\omega }^{\dagger }}$. In general case
$\Delta $ depends on the direction of the vector ${\bf v }_{F}$
and is determined by the self-consistency equation
\begin{equation}
\Delta ({\bf v}_{F},{\bf r})=2\pi N(0)T\sum\limits_{\omega
>0}<V_{{\bf v}
_{_{F}}{\bf v}_{F}^{\prime }}f_{\omega }({\bf v}_{F}^{\prime
},{\bf r)>}_{{\bf \theta }},  \label{sc}
\end{equation}
where $V_{{\bf v}_{_{F}}{\bf v}_{F}^{\prime }}$ is the interaction
potential. In our calculations we will consider two-dimensional
case; $N(0)= \frac{m}{2\pi }$ is 2D density of states and
$<...>_{{\bf \theta } }=\int\limits_{0}^{2\pi }\frac{d\theta
}{2\pi }...$ is the averaging over directions of 2D vector ${\bf
v}_{F}=(v_{F}\cos \theta ,v_{F}\sin \theta ).$ In general it is
possible to obtain a mixture of different symmetries of the order
parameter, $\Delta (\theta )=\Delta _{x^{2}-y^{2}}(\theta )+\Delta
_{xy}(\theta )+\Delta _{s}$ where $\Delta _{x^{2}-y^{2}}(\theta
)=\Delta _{1}\cos 2\theta ,$ $\Delta _{xy}(\theta )=\Delta
_{2}\sin 2\theta $ and $\Delta _{s}$ are the $d_{x^{2}-y^{2}}$,
$d_{xy},$ and the $s$-wave components of the order parameter
respectively. The corresponding interaction potential, $V_{\theta
\theta ^{\prime }}=V_{d1}\cos 2\theta \cos 2\theta ^{\prime
}+V_{d2}\sin 2\theta \sin 2\theta ^{\prime }+V_{s},$ must be
substituted in the self-consistency equation (\ref{sc}) for the
order parameter in each channel. The current density ${\bf j(r)}$
is found from the solution of the matrix equation (\ref{eil}) as
\begin{equation}
{\bf j(r)}=-4\pi ieN(0)T\sum\limits_{\omega >0}<{\bf
v}_{F}g_{\omega }({\bf v}_{F},{\bf r)>}_{\theta }.  \label{j}
\end{equation}

\begin{figure}[t]
\epsfysize 4.2cm \epsfbox[50 270 600 530]{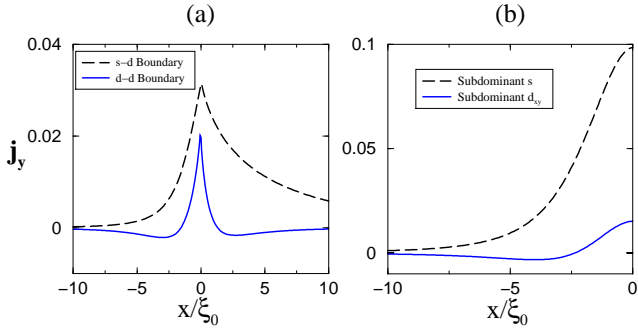}
\caption{\protect\small (a) Spontaneous current for $d$-$d$ and
$s$-$d$ junctions. The boundary is located at $x=0$. Calculations
are done at $t\equiv T/T_{c}=0.05$, with $T_{c2}=0.05T_{c}$ for
the $d$-$d$ case and $T_{cs}=0.1T_{c}$ and $T_{c2}=0.05T_{c}$ for
the $s$-$d$ junction. (b) Spontaneous current at the (110)-
surface of a $d$-wave superconductor at $t=0.05$ with
$T_{c2}=T_{cs}=0.1T_{c}$ for both $s$ or $d_{xy}$ subdominant
order parameters.} \label{jy-x}
\end{figure}

Here we consider three cases, (i) boundary between two
semi-infinite d-wave superconductors with crystalographic
orientations with respect to the boundary given by angles $\chi
_{l}$ and $\chi _{r}$ (``$d$-$d$ interface''), (ii) boundary
between an s-wave and a d-wave superconductor with $45^\circ$
orientation (``$45^\circ$ $s$-$d$ interface''), and (iii)
(110)-surface of a $d$-wave superconductor. In all three cases it
is possible to have time reversal symmetry breaking ground state.
The direction and magnitude of the spontaneous current depends on
the relative phases of the order parameters.

\begin{figure}[t]
\epsfysize 4.3cm \epsfbox[50 260 400 530]{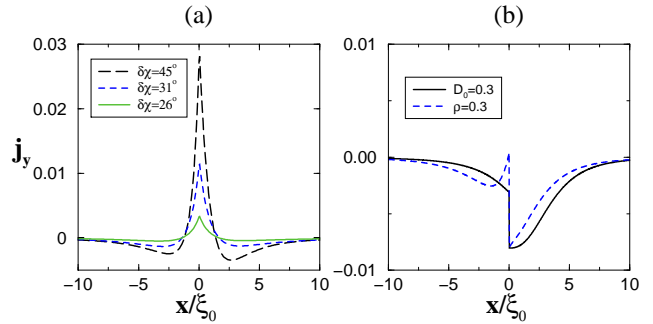}
\caption{{\protect\small (a) Spontaneous currents for
($0^\circ$-$\delta \chi$) junctions with different misorientation
angles. \ (b) Spontaneous current at imperfect
($0^\circ$-$45^\circ$) junctions. Solid line: junction with
transparency $D_{0}=0.3$. Dashed line: junction with roughness
$\rho=0.3$. All calculations are done at $t=0.1$.}} \label{jyd-x}
\end{figure}

Assuming constant order parameters on both sides of an interface, one can
obtain an analytical (non-selfconsistent) expression for the current density
\begin{eqnarray}
{\bf j}(x) &=&4\pi eN(0)T\sin \phi \label{jJ}  \\
&&\sum\limits_{\omega >0}\langle \frac{{\bf v}_{F}\Delta
_{l}\Delta _{r}{\rm sign}(\cos \theta )}{\Omega _{l}\Omega
_{r}+\omega ^{2}+\Delta _{l}\Delta _{r}\cos \phi }e^{-2|x|\Omega
_{r}/\left| v_{F}\cos \theta \right| }\rangle _{\theta },\nonumber
\end{eqnarray}
where $l$ ($r)$ labels left (right) side of the interface, and
$\Omega _{i}= \sqrt{\omega ^{2}+|\Delta _{i}|^{2}}$. This
expression is valid for arbitrary symmetry of the order
parameters $\Delta _{l,r}$. For a $d$-$d$ interface we have
$\Delta _{l}=\Delta _{0}(T)\cos 2(\theta -\chi _{l})$ and $\Delta
_{r}=\Delta _{0}(T)\cos 2(\theta -\chi _{r})$, where $\Delta
_{0}(T)$ depends on the superconducting coupling and temperature.

Our numerical calculations were based on Schopohl-Maki
parameterization of Green's functions \cite {schopohl} ,
\[
g={\frac{1-aa^{\dagger }}{1+aa^{\dagger }}},\qquad
f={\frac{2a}{1+aa^{\dagger }}},\qquad f^{\dagger
}={\frac{2a^{\dagger }}{1+aa^{\dagger }}},
\]
which transforms Eq. (\ref{eil}) into
\begin{eqnarray}
{\bf v}_{F}\cdot \nabla a &=&2\omega a-\Delta ^{\ast }a^{2}+\Delta  \label{a}
\\
-{\bf v}_{F}\cdot \nabla a^{\dagger } &=&2\omega a^{\dagger
}-\Delta a^{\dagger 2}+\Delta ^{\ast }.  \label{ad}
\end{eqnarray}
For positive $v_{x}$, Eq. (\ref{a}) (Eq. (\ref{ad})) is stable if
the boundary condition at $x\rightarrow -\infty $ ($+\infty $) is
chosen. The opposite is true for negative $v_{x}$. We use the
solutions in a homogeneous system, $a= \Delta / (\omega +\Omega
)$ and $a^{\dagger }= \Delta ^{\ast }/(\omega +\Omega)$ as
boundary conditions at $\pm \infty$. The values of $a$
($a^{\dagger}$) at all other points on the trajectory are then
easily found. The self-consistency is introduced through
iterations, assuming a constant order parameter in either half of
the junction for the first iteration.

Fig.\ \ref{jy-x}$a$ shows the spatial distribution of the
spontaneous current in $d$-$d$ and $s$-$d$ junctions. The left
superconuctor is a $d$-wave superconductor with $45^\circ$ crystal
orientation with respect to the boundary. The right side is
either an $s$-wave or a $d$-wave superconductor aligned with the
boundary. The current distribution is qualitatively different in
$s$-$d$ and $d$-$d$ junctions. In the $d$-$d$ case, the current
density is at maximum in a layer of width of about coherence
length, $\xi_0 =v_{F}/\pi \Delta ,$ along the boundary; there also
exists a counter-flow, spread over about $10\xi_0 $ on either side
of the boundary. The total current in $y$-direction is zero
within the numerical accuracy, {\em independently} on the right
and left sides of the junction. This effect can be called
``superscreening'', since the resulting magnetic field of the
spontaneous current is cancelled on the scale of $\sim 10\xi_0 \ll
\lambda _{L},\lambda _{J}$, the London and Josephson magnetic
penetration depths. Note that this has nothing to do with the
Meissner screening; it appears without taking into account the
vector potential of the magnetic field of the current (and makes
it unnecessary)\cite{MISSED}. On the contrary, in the $s$-$d$
junction the counter-flows are absent (unless the Meissner effect
is taken into account\cite{Huck}).

The same situation takes place near the surface, {\em if }the
subdominant pairing is present. Fig. \ref{jy-x}$b$ shows the
current distribution at the (110)-surface of a $d$-wave
superconductor. If $d_{xy}$ is the leading subdominant order
parameter, the form of the current distribution is similar to the
one in the $d$-$d$ boundary. The superscreening is absent if the
subdominant order parameter is $s$-wave.

\begin{figure}[t]
\epsfysize 4.5cm \epsfbox[50 260 400 530]{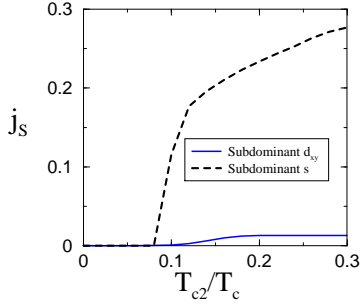}
\caption{{\protect\small The effect of subdominant interaction on
the spontaneous current $j_{S}$ ($\equiv j_{y}(x=0)$) at the
surface of a $d$ wave superconductor. A second order phase
transition happens at $T_{c2}=T=0.1T_{c}$}} \label{jydv}
\end{figure}

The superscreening effect can be obtained analytically from the
non-selfconsistent expression (\ref{jJ}) in case of $0$-$45^{\circ }$
junction. The nullification of the total current results from integrating
the spontaneous current
\[
\int_{0}^{\infty }dxj_{y}(x)\propto \langle \frac{\Delta _{l}\Delta _{r}\sin
\theta \ {\rm sign}(\cos \theta )}{\Omega _{l}\Omega _{r}+\omega ^{2}+\Delta
_{l}\Delta _{r}\cos \phi }\frac{v_{F}|\cos \theta |}{\Omega _{r}}\rangle
_{\theta }\sin \phi
\]
which is zero after angle averaging. Our numerical calculations however show
that in clean boundary junctions the total current is zero (within the
numerical accuracy) even after self-consistent calculation and at all other
misorientation angles (see Fig. \ref{jyd-x}$a$).

To understand the situation, let us recall that in a system with local
magnetic moment density ${\bf m(r)}$ the ``molecular currents'' flow with
density ${\bf j(r)=}c{\bf \nabla \times m(r).}$ In a superconductor with
order parameter $d_{x^{2}-y^{2}}+e^{i\phi _{0}}d_{xy}$ local
orbital/magnetic moment density
\begin{eqnarray}
{\bf m(r)} &\propto& \widehat{{\bf z}}\int_{0}^{2\pi
}\frac{d\theta }{2\pi }\left( \Delta _{1}(x)\cos 2\theta +\Delta
_{2}(x)e^{-i\phi _{0}}\sin 2\theta \right) \nonumber \\
&&\times \frac{1}{i}\frac{\partial }{\partial \theta }\left( \Delta
_{1}(x)\cos 2\theta +\Delta _{2}(x)e^{i\phi _{0}}\sin 2\theta \right)
\nonumber \\
&=&2\Delta _{1}(x)\Delta _{2}(x)\widehat{{\bf z}}\sin \phi _{0}
\nonumber.
\end{eqnarray}
The contribution to the spontaneous current is thus ${\bf
j(r)\propto \nabla \times m(r)\parallel \hat{y}}$. Notice that
the same expression is obtained from the Ginzburg-Landau equations
\cite{bailey} (${\bf j} \propto \nabla \times (\widehat{z}\ {\rm
Im}\ d_1({\bf r}) d_2({\bf r})^*)$). The total current in
$y$-direction due to this mechanism is $I_{\rm total}\propto
\int_{\Omega} d{\bf S\cdot \nabla \times m} =\oint_{\partial
\Omega} d{\bf l \cdot m}=0$, where $\Omega$ is a cross section
perpendicular to the junction from $x=-\infty$ to $\infty$ and
$\partial \Omega$ is its boundary. The latter integral is
obviously zero because $d{\bf l \cdot m}=0$ (${\bf m}\parallel
\hat{\bf z}$) everywhere except where the contour closes ($x= \pm
\infty$), but there ${\bf m}=0$. This is certainly not the case
in $s$-$d$ junctions (cf. Fig. \ref{jy-x}). (Of course, since the
Meissner currents must be taken into account in this case, the
results presented in Fig. \ref{jy-x} are valid only if the system
size is much less than the London penetration depth.)

We also calculate the spontaneous current for an imperfect
boundary, i.e a boundary with arbitrary transparency $D_{0}$ and
also with finite roughness $\rho$. We use Zaitsev's boundary
condition \cite{Zaitsev,yip2} to incorporate the finite
transparency effect. For surface roughness we assume a thin layer
with scattering centers at the junction \cite{golubov}. We take
the mean free path $l$ and the layer thickness $d$ to zero while
keeping $\rho\equiv d/l$ finite. The details of calculations will
be given in a separate publication. Here we only present the
results of our calculation for asymmetric ($0^{\circ
}$-$45^{\circ }$) $d$-$d$ junction in Fig. \ref{jyd-x}$b$. As is
clearly seen, the spontaneous currents now does not necessarily
have a counterflow (at small $\rho$ or $D_0 \approx 1$ there will
be some counterflow), and the exact supersreening no longer takes
place. They are now carried merely by Andreev bound states at the
interface, the same as in $s$-$d$ or SND \cite{Huck} junctions.
\begin{figure}[t]
\epsfysize 4.5cm \epsfbox[50 260 400 530]{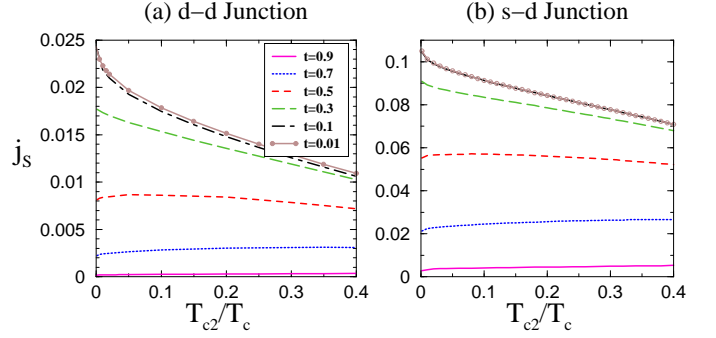}
\caption{{\protect\small Suppression of spontaneous current by
subdominant order parameter.\ (a) A $d$-$d$ grain boundary. \ (b)
An $s$-$d$ interface. In the $s$-$d$ case, we have taken the same
$T_c$ for both sides and also $T_{cs}=T_{c2}$.}} \label{jy-Tc2}
\end{figure}
Although near realistic surfaces/interfaces with $d$-$d$ ordering
the superscreening is not complete, the magnetic fields created
by the spontaneous currents are nevertheless suppressed on very
short distances. This can be practically important for attempting
to build a ``quiet'' qubit based on such
junctions\cite{Feigel'man!}.

Fig.\ \ref{jydv} presents the spontaneous current as a function of the
subdominant critical temperature $T_{c2}$ at the (110)-surface of a d-wave
suprconductor. One notices that the spontaneous current vanishes when $%
T_{c2} < T$. In fact, at temperatures below $T=T_{c2}$ the
subdominant order parameter starts to appear at the surface
through a second-order phase transition. Spontaneous symmetry
breaking and generation of the spontaneous current are the
consequences of the emergence of this second order parameter. The
symmetry of the subdominant order parameter is dictated by
whichever channel ($s$ or $d_{xy}$) has stronger interaction
potential.

In the $d$-$d$ and $s$-$d$ interfaces on the other hand, the
subdominant order parameter is induced by the proximity to a
different superconductor. One important difference is that unlike
the surface case, at the $d$-$d$ or $s$-$d$ interfaces the
presence of the subdominant order parameter is not necessary for
generation of spontaneous current. From Eqs. (\ref{j}) and
(\ref{eil}) we see that it is the Green's function (the pairing
{\em amplitude}), not the order parameter (pairing {\em
potential}), which determines the current. In fact, the presence
of subdominant order parameter does not always increase the
spontaneous current. At low temperatures, it actually {\em
decreases} the spontaneous current \cite{tanaka}. This
counter-intuitive effect is displayed in Fig. \ref {jy-Tc2} in
which the spontaneous current is plotted as a function of
$T_{c2}$. The temperature used in the calculations is $t=0.1$ and
we take the same $T_{c}$ for both $d$ and $s$-wave
superconductors. Increasing $T_{c2}$ increases the interaction in
the subdominant channel and therefore the magnitude of the
subdominant order parameter. The spontaneous current on the other
hand decreases with increasing $T_{c2}$. The situation is the
same for both $d$-$d$ and $s$-$d$ interfaces.

\begin{figure}[t]
\epsfysize 4.3cm \epsfbox[-400 10 400 700]{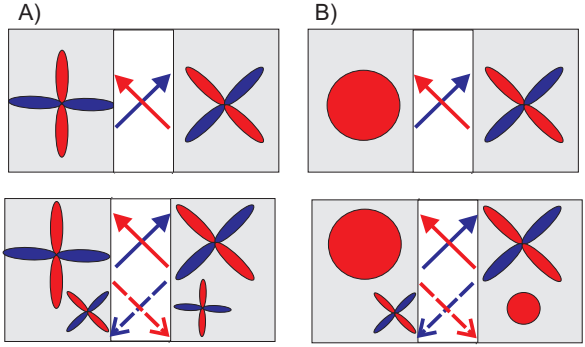}
\caption{\protect\small The DND and SND model of a ${\cal
T}$-breaking junction.\ (a) DND junction. The normal region
contains current-carrying Andreev bound states (arrows); in
equilibrium net current across the boundary is zero,while the
spontaneous currents flow along the normal layer (above). If
subdominant order parameter is present, the additional set of
Andreev levels in equilibrium carries spontaneous current in the
opposite direction (below). The model gives the same predictions
for SND case (b). } \label{andrev}
\end{figure}

The decrease of the spontaneous currents when there is
interaction in the subdominant channel may seem paradoxical.
Nevertheless it is easy to understand in the DND-model of ${\cal
T}$-breaking junction. \cite{Huck,Zagoskin} (Fig. \ref{andrev}).
First consider the case without subdominant order parameters. The
spontaneous currents in this model flow exclusively within the
normal layer and are carried by ``zero'' and ``$\pi $''- Andreev
bound states, which connect the lobes of d-wave order parameter
with the same and opposite signs respectively: in equilibrium
there is no net current across the boundary. Now let us assume
that the subdominant order parameters are present. Due to
continuity, they must have the same phase as the dominant order
parameter on the other side (Fig. \ref{andrev}). Therefore now we
will have two extra sets of current-carrying Andreev states, the
ones linking the {\em subdominant} order parameters, and it is
obvious that the spontaneous currents they carry will always flow
opposite to the currents carried by the ``dominant-dominant''
states.

In conclusion, we have investigated the spontaneous currents near
the surface and $d$-$d$ and $s$-$d$ boundaries in $d$-wave
superconductors. We obtained the contributions to the spontaneous
currents due to the proximity effect and due to the subdominant
order parameter generation, and found that at interfaces the
latter generally decreases the magnitude of the effect. In $d$-$d$
junctions, we separated the contribution from the local
orbital/magnetic moment of the condensate; this contribution
dominates spontaneous currents in clean $d$-$d$ junctions, which
explains the superscreening of the spontaneous currents in such
systems.

We would like to thank S. Rashkeev, G. Rose, A. Smirnov and I.
Herbut for helpful discussions.


\begin{references}
\bibitem{MISSED} A similar distribution of spontaneous current density on
(110) surfaces and interfaces of d-superconductors was obtained
in \cite {FogelstromYip} (Fig.4a).
\end{references}
\end{document}